\RequirePackage{fixltx2e}
\documentclass[aip,jcp,reprint,floatfix]{revtex4-1}

\usepackage{amsmath,amsfonts} %
\usepackage{wasysym} %
\usepackage{graphicx} %
\usepackage[dvipsnames,table]{xcolor} %
\usepackage[acronym]{glossaries} %
\usepackage{dcolumn} %
\usepackage{bm} %
\usepackage{ifpdf} %
\usepackage{braket} %
\usepackage{siunitx} %
\usepackage{fancyhdr} %
\usepackage{microtype} %
\usepackage{etoolbox} %
\usepackage{natbib} %
\usepackage[byname]{smartref} %
\usepackage[breaklinks, %
colorlinks, linkcolor=Blue, citecolor=Blue, urlcolor=Blue]{hyperref} %
\usepackage[version=3]{mhchem}
\usepackage{multirow}
\AtBeginDocument{\usepackage{booktabs}} 

\graphicspath{{pics/}}

\newcommand{\bfr}{\mathbf{r}}

\hypersetup{ %
  pdftitle={Fast numerical evaluation of time-derivative nonadiabatic
    couplings for mixed quantum-classical methods}, %
  pdfauthor={Ilya G. Ryabinkin, Jayashree Nagesh, and Artur F.
    Izmaylov}, %
  pdfsubject={Nonadiabatic dynamics}, %
  pdfkeywords={} %
}

\newacronym{NAC}{NAC}{nonadiabatic coupling} %
\newacronym{TDNAC}{TDNAC}{time-derivative nonadiabatic coupling} %
\newacronym{TDSE}{TDSE}{time-dependent Schr\"odinger equation} %
\newacronym{SH}{SH}{surface hopping} %
\newacronym{CI}{CI}{conical intersection} %
\newacronym{MR-CI}{MR-CI}{multi-reference configuration interaction} %
\newacronym{CIS}{CIS}{configuration interaction singles} %
\newacronym{MCSCF}{MCSCF}{multi-configurational self-consistent field}
\newacronym{PES}{PES}{potential energy surface} %
\newacronym{FC}{FC}{Franck--Condon} %
\newacronym{GP}{GP}{geometric phase} %
\newglossaryentry{RDM}{%
  type = \acronymtype, %
  name = {RDM}, %
  description = {reduced density matrix}, %
  text = {RDM}, %
  first = {reduced density matrix (RDM)}, %
  firstplural = {reduced density matrices (RDMs)}, %
  plural = {RDMs}} %

\begin{document}

\title{Fast Numerical Evaluation of Time-Derivative Nonadiabatic
  Couplings for Mixed Quantum-Classical Methods}

\author{Ilya G. Ryabinkin} %
\affiliation{Department of Physical and Environmental Sciences,
  University of Toronto Scarborough, Toronto, Ontario, M1C\,1A4,
  Canada} %
\affiliation{Chemical Physics Theory Group, Department of Chemistry,
  University of Toronto, Toronto, Ontario M5S\,3H6, Canada} %

\author{Jayashree Nagesh} %
\affiliation{Chemical Physics Theory Group, Department of Chemistry,
  University of Toronto, Toronto, Ontario M5S\,3H6, Canada} %

\author{Artur F. Izmaylov} %
\affiliation{Department of Physical and Environmental Sciences,
  University of Toronto Scarborough, Toronto, Ontario, M1C\,1A4,
  Canada} %
\affiliation{Chemical Physics Theory Group, Department of Chemistry,
  University of Toronto, Toronto, Ontario M5S\,3H6, Canada} %

\date{\today}

\begin{abstract}
  We have developed a numerical differentiation scheme which
  eliminates evaluation of overlap determinants in calculating the
  \glspl{TDNAC}. Evaluation of these determinants was the bottleneck in
  previous implementations of mixed quantum-classical methods using
  numerical differentiation of electronic wave functions in the
  Slater-determinant representation. The central idea of our approach
  is, first, to reduce the analytic time derivatives of Slater
  determinants to time derivatives of molecular orbitals, and then to
  apply a finite-difference formula. Benchmark calculations prove the
  efficiency of the proposed scheme showing impressive
  several-order-of-magnitude speedups of the \gls{TDNAC} calculation
  step for midsize molecules.
\end{abstract}

\glsresetall

\pacs{}

\maketitle


Simple quantum-classical methods for simulating nonadiabatic
dynamics, Ehrenfest and fewest-switches surface hopping
(FSSH),\cite{Tully:1990/jcp/1061, Tully:1998/fd/407} often provide
accurate and efficient ways to investigate chemical processes
involving several electronic states. The simplicity of these methods
stems from restricting quantum mechanical consideration to the
electronic part and treating the nuclear part classically with minimal
intervention of quantum mechanics. For describing the quantum evolution of
the electronic subsystem its non-stationary electronic wave function is
written in terms of the adiabatic eigenfunctions
$\{\Psi_J(\bfr;\mathbf{R})\}$ of the electronic Hamiltonian $\hat H_e$
as
\begin{equation}
  \label{eq:exp}
  \psi(t, \bfr; \mathbf{R}(t)) = \sum_J c_J(t)
  \Psi_J(\bfr;\mathbf{R}(t)).
\end{equation}
The time-dependent coefficients $c_J(t)$ then can be obtained via
projecting the time-dependent electronic Schr\"odinger equation onto
the orthonormal basis of $\{\Psi_J(\bfr;\mathbf{R})\}$ 
  (atomic units are assumed hereinafter)
\begin{equation}
  \label{eq:SH_elec}
  i \frac{\mathrm{d}c_K}{\mathrm{d}t} = \sum_J c_J
  \big(\delta_{KJ}E_J(\mathbf{R}) - i\tau_{KJ}\big),
\end{equation}
where $\delta_{KJ}$ is the Kronecker delta, $E_J(\mathbf{R})$ are
the adiabatic potential energy surfaces (PESs), and
\begin{equation}
  \label{eq:tdnacs}
  \tau_{KJ} = \Braket{\Psi_K | \partial_t\Psi_J}, \quad K \ne J,
\end{equation}
are the \glspl{TDNAC}. Using the chain rule one can further decompose
$\tau_{KJ}= \dot{\mathbf{R}}\cdot \mathbf{d}_{KJ}$, where
$\mathbf{d}_{KJ} = \Braket{\Psi_K | \nabla_\mathbf{R}\Psi_J}$ is the
$3M$-dimensional ($M$ is the number of nuclei in the system)
derivative couplings vector, and $\dot{\mathbf{R}}$ is the
$3M$-dimensional nuclear velocity vector. $\mathbf{d}_{KJ}$ are
implemented analytically for some electronic structure methods, such
as \gls{MCSCF},\cite{gamessus, G09-short} \gls{CIS},\cite{jcp/135/234105/2011,Zhang:2014ci}  and \gls{MR-CI}.\cite{columbus_nacs_1, columbus_nacs_2} However, many electronic structure
methods either lack of the analytic implementation (\emph{e.g.}
XMCQDPT~\cite{Granovsky:2011/jcp/214113}) or have intrinsic problems
in their definition.\cite{Ou:2014/jcp/024114, Li:2014/jcp/244105,
  Li:2014/jcp/014110, Alguire:2015/jpcb/7140, Zhang:2015/jcp/064109,
  Ou:2015/jcp/064114} Of course, numerical differentiation is always
an option for evaluation of $\mathbf{d}_{KJ}$'s but it is also quite
computationally expensive considering the dimensionality of these
quantities.
 
On the other hand, it has been recognized that it is more efficient to
apply numerical differentiation to \glspl{TDNAC}
directly.\cite{Hammes:1994/jcp/4657} For example, any of the following
formulae can be employed
\begin{align}
  \label{eq:fd_scheme_1st}
  \tau_{KJ} = {} & \frac{1}{\Delta t}\Braket{\Psi_K(t)|\Psi_J(t+\Delta
    t)} + o(\Delta t), \\
  \label{eq:fd_scheme_2nd}
  \tau_{KJ} = {} & \frac{1}{4\Delta t}\Big(\Braket{\Psi_K\left(t -
      \Delta
      t\right)|\Psi_J(t + \Delta t)}  \nonumber \\
  & \ \quad -\Braket{\Psi_K(t + \Delta t)|\Psi_J(t - \Delta t)}\Big) +
  o(\Delta t^2),
\end{align}
giving rise to the first-order forward and the second-order central
finite difference schemes, respectively. 
Moreover, as was shown in Ref.~\onlinecite{Meek:2014/jpcl/2351}, use of numerical
  \glspl{TDNAC} can be advantageous close to surface crossings. 
However, in the conventional
FSSH method, $\mathbf{d}_{KJ}$ quantities are also needed to rescale
nuclear velocities if a hop between electronic surfaces
$E_{K}(\mathbf{R})$ and $E_{J}(\mathbf{R})$ occurs.

Recently, to avoid numerical evaluation of $\mathbf{d}_{KJ}$, a
simpler version of the FSSH method has been
suggested.\cite{Tapavicza:2007/prl/023001} In this simplified version
nuclear velocities are rescaled uniformly after a surface hop. It was
shown that this simplified scheme can adequately model nonadiabatic
dynamics and deviates from the regular FSSH algorithm only in regions
where $\mathbf{d}_{KJ}$'s change rapidly, but these deviations have
only a minor effect on overall dynamics.\cite{Pittner:2009/cp/147}
Thus, if we focus on the simplified FSSH method, the only required nonadiabatic coupling terms will be
\glspl{TDNAC}.

In commonly used numerical formulations [Eq.~\eqref{eq:fd_scheme_1st}
or \eqref{eq:fd_scheme_2nd}], \glspl{TDNAC} are obtained from an electronic overlap matrix
  \begin{equation}
    \label{eq:cis_overlap}
    \Sigma_{KJ}(t, t+\Delta t) = \Braket{\Psi_K(t)|\Psi_J(t + \Delta
      t)},
  \end{equation}
  whose matrix elements require Slater-determinant pair overlaps for
electronic wave functions in the Slater-determinant
representation~\cite{Pittner:2009/cp/147} 
  \begin{eqnarray}
    \notag \Sigma_{KJ}(t, t+\Delta t) & = & \sum_{\{p,q\}}
    C_{\{p\}}^K(t) C_{\{q\}}^J(t + \Delta t) \\ 
    \label{eq:Pitt} 
    & & \times\braket{\Phi_{\{p\}}(t)|\Phi_{\{q\}}(t +
      \Delta t)}, 
  \end{eqnarray}
where $C_{\{p\}}^{K}$ and $C_{\{q\}}^{J}$ are coefficients of the
$\ket{\Phi_{\{p\}}(t)}$ and $\ket{\Phi_{\{q\}}(t+\Delta t)}$ Slater
determinants. Here, we use collective indices $\{p\}$ and $\{q\}$
denoting sets of orbitals present in the Slater determinants
$\ket{\Phi_{\{p\}}(t)}$ and $\ket{\Phi_{\{q\}}(t+\Delta t)}$. This
scheme quickly becomes computationally expensive with the system size,
because it requires evaluation of many Slater-determinant pair
overlaps $\braket{\Phi_{\{p\}}(t)|\Phi_{\{q\}}(t + \Delta t)}$ given
by the L\"{o}wdin formula~\cite{Lowdin:1955/pr/1474, Mayer:2003/BOOK}
\begin{equation}
  \label{eq:lowd_f}
  \braket{\Phi_{\{p\}}(t)|\Phi_{\{q\}}(t + \Delta t)} =
  \det{\mathbf{S}[\{p\}\{q\}]},
\end{equation}
where $\mathbf{S}[\{p\}\{q\}]$ is the overlap matrix of molecular
orbitals comprising the determinants $\ket{\Phi_{\{p\}}(t)}$ and
$\ket{\Phi_{\{q\}}(t + \Delta t)}$. The L\"{o}wdin formula appears as
a result of non-orthogonality between sets of orbitals at different
times. The computational cost of $\det{\mathbf{S}[\{p\}\{q\}]}$
calculation grows cubically with the number of electrons in the
system, $N_e$.\cite{nr} Considering that all pairs of 
determinants in Eq.~\eqref{eq:Pitt} need to be evaluated, use of 
Eq.~\eqref{eq:Pitt} in large systems makes the evaluation of
\glspl{TDNAC} the bottleneck of mixed quantum-classical simulations.
  
In this Letter we show how computing of the determinant
overlaps can be avoided in numerical evaluation of \glspl{TDNAC}
without introducing any approximations and by making the procedure
faster by at least a factor of $\sim N_\text{occ}^3$ 
for each determinant pair ($N_\text{occ} = N_e/2$ for the closed-shell case). 
We illustrate the performance of our approach by computing \glspl{TDNAC} at the
\gls{CIS} level of theory, which is one of the simplest methods for
treating excited states. Our developments can be straightforwardly
applied to any other method presenting wave functions as linear
combinations of Slater determinants (e.g., MR-CI or MCSCF).

In the \gls{CIS} method, excited-state wave functions are written as
linear combinations of singly-excited Slater
determinants $\ket{\Phi_i^a} = \hat a_a^\dagger \hat a_i \ket{\Phi_0}$
obtained from the ground-state Hartree--Fock determinant
$\ket{\Phi_0}$
\begin{equation}
  \label{eq:cis_wf}
  \ket{\Psi_K} = \sum_{ia} C_{ia}^K \ket{\Phi_i^a},
\end{equation}
with coefficients $C_{ia}^K$ defined by the secular matrix problem
$\hat{\mathbf{H}}_e\mathbf{C} =\mathbf{E}\mathbf{C}$. Here, we follow
the common convention where subscripts $a,b,c, \ldots$ denote virtual
orbitals, $i,j,k, \ldots$ label occupied orbitals, and $p,q,r \ldots$
are used for any type of orbitals.

To avoid computational difficulties associated with overlap
determinants [Eq.~\eqref{eq:lowd_f}] we will start with the formal
definition of \glspl{TDNAC} as time derivatives
[Eq.~\eqref{eq:tdnacs}] and postpone applying a finite difference
scheme until we account for the anti-symmetric structure of Slater
determinants. Assuming real-valued molecular orbitals and \gls{CIS}
coefficients, \glspl{TDNAC} can be written as 
  \begin{align}
    \label{eq:nacs_cis}
    \tau_{KJ} = {} & \sum_{ijab} \Bigl( C_{ia}^K \partial_t C_{jb}^J
    \Braket{\Phi_i^a|\Phi_j^b} + C_{ia}^K C_{jb}^J
    \Braket{\Phi_i^a|\partial_t\Phi_j^b} \Bigr).
  \end{align}
  All terms in Eq.~\eqref{eq:nacs_cis} refer to the same $t$, hence,
  there are no complications with orbital non-orthogonality as
  in Eq.~\eqref{eq:Pitt}. One may apply the Slater--Condon rules to
  the first term, but not to the second one, since the time derivative
  $\partial_t$ is \emph{not} an operator in the space of electronic
  variables. Instead, we differentiate determinants $\ket{\Phi_j^b}$
  directly
  \begin{equation}
    \label{eq:det_der}
    \partial_t \ket{\Phi_j^b} = \sum_{k \ne j} \ket{\Phi_{jk}^{bk'}} +
    \ket{\Phi_j^{b'}}, 
  \end{equation}
  where the notation $\ket{\Phi_{p}^{q'}}$ means that a molecular
  orbital $\phi_p$ is replaced with the time derivative $\partial_t
  \phi_q$. Therefore, \glspl{TDNAC} between determinants become
  \begin{equation}
    \label{eq:ovpl}
    \Braket{\Phi_i^a|\partial_t\Phi_j^b} = \sum_{k \ne j}
    \braket{\Phi_i^a|\Phi_{jk}^{bk'}} + \braket{\Phi_i^a|\Phi_{j}^{b'}}.
  \end{equation}
  The last term in Eq.~\eqref{eq:ovpl} is reduced to $\delta_{ij}
  \braket{\phi_a|\partial_t \phi_b}$, while only one term with $k = i$
  and $a = b$ from the sum over $k$ survives because of orthogonality
  conditions: $\braket{\phi_p|\partial_t \phi_p} = 0$ (for real
  orbitals) and $\braket{\phi_p|\phi_q} = \delta_{pq}$. Finally, we
  have:
\begin{equation}
  \label{eq:ovpl_fin}
  \Braket{\Phi_i^a|\partial_t \Phi_j^b} = \delta_{ij}
  \Braket{\phi_a|\partial_t \phi_b} - P_{ij}\delta_{ab}
  \Braket{\phi_j|\partial_t \phi_i},
\end{equation}
where $P_{ij}$ is an additional phase factor which depends on the
ordering convention for the orbitals in the Slater
  determinants. There are two common choices which
  result in different $P_{ij}$ values:
\begin{subequations}
  \begin{align}
    \label{eq:det_sign_conv_1}
    \ket{\Phi_i^a} & = \det{\left\lbrace \ldots, \phi_{i-1},
        \phi_a ,\phi_{i+1}, \ldots \right\rbrace}, \ P_{ij} = 1, \\
    \label{eq:det_sign_conv_2}
    \ket{\Phi_i^a} & = \det{\left\lbrace \ldots, \phi_{i-1},
        \phi_{i+1}, \ldots , \phi_a\right\rbrace}, \ P_{ij} =
    (-1)^{|j-i|},
  \end{align}
\end{subequations}

Use of Eq.~\eqref{eq:ovpl_fin} leads to substantial reduction of
\gls{TDNAC} computation scaling because Eq.~\eqref{eq:nacs_cis} is
simplified to
\begin{align}
  \label{eq:nacs_cis_simpl}
  \tau_{KJ} = {} & \sum_{ia} C_{ia}^K
  \partial_t C_{ia}^J + \sum_{iab} C_{ia}^K C_{ib}^J
  \Braket{\phi_a|\partial_t \phi_b} \nonumber
  \\
  - &\sum_{ija} P_{ij} C_{ia}^K C_{ja}^J
  \Braket{\phi_j|\partial_t\phi_i}.
\end{align}
Each term in Eq.~\eqref{eq:nacs_cis_simpl} scales as
$N_\text{occ}N_\text{virt}$, $N_\text{occ}N_\text{virt}^2$, and
$N_\text{occ}^2N_\text{virt}$, respectively, where $N_{\rm occ}$ and
$N_{\rm virt}$ are the numbers of occupied and virtual orbitals. 
 In typical calculations, 
$N_{\rm virt}>N_{\rm occ}$, and the second term [Eq.~\eqref{eq:nacs_cis_simpl}] is dominating in 
the overall computational cost providing the overall cubic scaling with the size of the system. 
This scaling should be compared with the $N_\text{occ}^5N_\text{virt}^2$
scaling of Eq.~\eqref{eq:Pitt} for the CIS method with the determinant scheme. Note that
evaluation of all terms in Eq.~\eqref{eq:nacs_cis_simpl} can be reformulated as highly efficient
matrix-matrix multiplications.

The possibility of reducing time derivatives of determinants to time
derivatives of orbitals has been already mentioned in several
works.\cite{Shenvi:2009/jcp/174107, Akimov:2013/jctc/4959} However,
Eq.~\eqref{eq:ovpl_fin} has never been derived explicitly; our
treatment, therefore, provides a rigorous foundation for the orbital formulation 
 and for its extension to multi Slater determinant wave functions.

  To apply Eq.~\eqref{eq:ovpl_fin}, one has to convert it into a
  corresponding finite-difference expression. Any finite-difference
  expression requires continuity of orbitals at different times.
  However, orbital phases at different times are
  \textit{arbitrary}, reflecting the existence of the wave function gauge (phase) degree of freedom. 
  Thus an appropriate orbital phase matching and tracking procedure is necessary.

  The finite-difference counterpart of Eq.~\eqref{eq:ovpl_fin} is
  obtained by substituting
  \begin{equation}
    \label{eq:orbital_fd}
    \Braket{\phi_p|\partial_t \phi_q} \to \frac{1}{\Delta
      t}S_{pq}(t,t+\Delta t), \quad p \ne q
  \end{equation}
  where
  \begin{equation}
    \label{eq:ovlp_diff}
    S_{pq}(t,t+\Delta t) = \braket{\phi_p(t)|\phi_q(t + \Delta t)}
  \end{equation}
  is the orbital overlap matrix. 
  To keep track of relative signs of orbitals at $t$
  and $t + \Delta t$ we introduce an integer-valued matrix
  $\mathbf{O}$, which is obtained from $\mathbf{S}(t, t+\Delta t)$ by
  rounding off its matrix elements to $\pm 1$ or $0$. $\mathbf{O}$ has
  a structure of the signed permutation matrix as long as $\Delta t$ is sufficiently small.
  \footnote{Large $\Delta t$'s will lead to appearance of zero columns and 
  rows in $\mathbf{O}$ and thus will be easy to detect.} 
  Performing the permutation and sign changes of molecular orbital coefficients in 
  $\mathbf{C}(t+\Delta t)$ according to $\mathbf{O}$ we obtain 
  matrix $\mathbf{\widetilde C}(t + \Delta t)$.
  The set of orbitals $\mathbf{\widetilde C}(t +\Delta t)$ is subsequently used to calculate the \gls{CIS}
  coefficients at the moment $t + \Delta t$. Tracking and phase
  matching for the \gls{CIS} states remain the same as for the
  determinant-based procedure.\cite{Pittner:2009/cp/147}
  Computational overhead for the orbital
  tracking and phase matching scales as $(N_{\rm occ}+N_{\rm virt})^2$ and 
  is negligible compare to other components of the algorithm.


To test the accuracy and efficiency of the proposed scheme, we benchmark it against
the conventional scheme based on Eqs.~\eqref{eq:Pitt} and
\eqref{eq:lowd_f} as implemented by Pittner and coworkers in the
\textsc{Newton-X} program.\cite{Barbatti:2007/jppa/228,
  Barbatti:2013/wcms/26} 
  Table~\ref{tab:accrcy}  shows that accuracies of the orbital- and determinant-based
  schemes are very similar as could be expected from numerical schemes of the same order. 
  For the efficiency comparison it is worth noting that 
  the \textsc{Newton-X} scheme has been used with a
  screening threshold of \num{5e-5} for the products of \gls{CIS}
  coefficients to reduce the number of determinant overlaps in
  Eq.~\eqref{eq:Pitt}.%
 Our implementation uses matrix-matrix multiplication and thus do not
employ a screening procedure. Table~\ref{tab:nacs_speedup} 
illustrates speedups achieved by the current scheme for two midsize
organic molecules and various basis sets. The speedups are especially
prominent for small basis sets, where $N_\text{occ}$ is comparable to
the total number of basis functions. Increase of the basis set size
makes the \gls{CIS} coefficient product screening more productive, but
the orbital-based scheme still outperforms the conventional scheme by
more than two orders of magnitude.
  \begin{table}
    \centering
    \caption{ Errors of numerical differentiation for representative \protect\glspl{TDNAC}
      in \ce{CH2NH2+} using the 3-21G basis.\cite{misc:CH2NH2} 
      \protect\gls{TDNAC} values converged up to $1\times10^{-7}$, $\tau_{31} = \num{-1.170e-3}$ and $\tau_{21} = \num{-2.375e-4}$ are obtained with 0.05 fs time-step.}
    \sisetup{
      table-format = -1.3e-1,
      table-number-alignment = center-decimal-marker,
      table-align-exponent = false,
    }
    \begin{tabular*}{1.0\linewidth}{
        @{\extracolsep{\fill}}l*{3}{S}@{}}
      \toprule
      & \multicolumn{1}{l}{Step size, \si{\femto\second}} & {Determinant} & {Orbital} \\
      \midrule
      \multirow{3}{*}{$\Delta\tau_{31}$} 
      & 0.1                             & -1.4E-06        &   -1.6E-06 \\
      & 0.2                             & -7.5E-06        &   -8.1E-06 \\
      & 0.5                             & -5.3E-05        &   -5.7E-05 \\
      \cmidrule{2-4}
      \multirow{3}{*}{$\Delta\tau_{21}$} 
      & 0.1                             &     -2.0E-07    &   <1.0E-07 \\
      & 0.2                             &     -7.0E-07    &   -2.0E-07 \\
      & 0.5                             &     -5.0E-06    &   -2.5E-06 \\
      \bottomrule
    \end{tabular*}
    \label{tab:accrcy}
  \end{table}
  \begin{table}[b]
    \centering
    \caption{{Relative speedups for a single time step evaluation of \protect\glspl{TDNAC} as
      compared to the \textsc{Newton-X} procedure. 
       In all calculations \protect\glspl{TDNAC} between the
      first three lowest electronic states were evaluated.}} 
    \label{tab:nacs_speedup}
    \begin{tabular*}{1.0\columnwidth}{
        @{\extracolsep{\fill}}lc*{3}{S[table-number-alignment=center]}@{}} %
      \toprule
      Molecule                     & Basis set & {$N_{\rm occ}$\footnotemark[1]} &
      {$N_{\rm virt}$}             & {Speedup} \\
      \midrule
      \ce{C18H14O}\footnotemark[2] & STO-3G    & 46        & 44  & 400       \\
      & 6-31G**   & 46        & 290 & 248       \\
      & cc-pVTZ   & 46        & 701 & 172       \\[0.5ex]
      \ce{C25H18}\footnotemark[3]  & STO-3G    & 59        & 59  & 1372      \\
      & 6-31G**   & 59        & 381 & 292       \\
      \bottomrule
    \end{tabular*}
    \footnotetext[1]{Excluding $1s$ core orbitals of \ce{C} and \ce{O}.}.
    \footnotetext[2]{4-(2-naphthylmethyl)-benzaldehyde.}
    \footnotetext[3]{9-((1-naphthyl)-methyl)-anthracene.}
  \end{table}

In practical applications one is more concerned with the total
simulation time. Apart from \glspl{TDNAC} calculations, mixed
quantum-classical nonadiabatic simulations include also the
electronic-structure and classical-dynamics steps. { 
For the electronic-structure CIS calculations we have used the Gaussian program.\cite{G09-short}} 
To give an idea of the overall speedup, we consider the first \SI{50}{\femto\second} of a
single FSSH trajectory for the \ce{C18H14O} molecule using the 6-31G**
basis set and the \SI{0.2}{\femto\second} time-step. {The electronic-structure 
part involved evaluation of characteristics for the three lowest electronic states.} 
On a single Intel
Xeon X5650 @ 2.67GHz CPU, it took \SI{99}{\hour} in total to complete
this trajectory for the original \textsc{Newton-X} procedure with
\SI{54}{\hour}\footnote{In fact, \textsc{Newton-X} spent twice as much
  as that (\SI{108}{\hour}) at \protect\glspl{TDNAC} calculations
  since it repeated the same calculations for both $\alpha$ and
  $\beta$ sets of molecular orbitals assuming the spin-unrestricted
  formalism. Here, we accounted only for a single set of calculations 
  necessary in the spin-restricted formulation.} spent on the
\gls{TDNAC} calculations, the corresponding numbers for our algorithm
are only \SI{45}{\hour} and \SI{12}{\minute}.


In conclusion, we have developed a numerical procedure which
eliminates evaluation of overlap determinants in \glspl{TDNAC} calculations.
. This elimination produces tremendous speedup in
quantum-classical nonadiabatic simulations where evaluation of
\glspl{TDNAC} was the bottleneck. The central idea of our approach is to
postpone introducing a finite-difference scheme,
[Eqs.~\eqref{eq:fd_scheme_1st} or \eqref{eq:fd_scheme_2nd}] and to
convert the expression for \glspl{TDNAC} given in terms of
many-electron wave functions, Eq.~\eqref{eq:nacs_cis}, into the
corresponding orbital-based formula, Eq.~\eqref{eq:nacs_cis_simpl}.
This alternation of steps allows us to manipulate with orthogonal
molecular orbitals and thus to avoid overlap determinants that
arose as a result of orbital non-orthogonality in the determinant
formulation. Benchmark calculations have proven the efficiency of the
proposed scheme and illustrated its potential for mixed
quantum-classical studies of medium and large molecules.

We thank A. Akimov for helpful discussions. A.F.I greatly
appreciates financial support by Alfred P. Sloan Foundation and
NSERC of Canada through the Discovery
Grants Program. J.N.'s work was supported by the U.S. Air Force Office
of Scientific Research under contract (to P. Brumer) number
FA9550-13-1-0005.


%

\end{document}